# Simple and Fast Polarization Tracking algorithm for Continuous-Variable Quantum Key Distribution System Using Orthogonal Pilot Tone

Yan Pan, Heng Wang, Yun Shao, Yaodi Pi, Ting Ye, Shuai Zhang, Yang Li, Wei Huang, and Bingjie Xu

*Abstract*—To reduce the influence of random channel polarization variation, especially fast polarization perturbation, for continuous-variable quantum key distribution (CV-QKD) systems, a simple and fast polarization tracking algorithm is proposed and experimentally demonstrated. The algorithm is implemented by an orthogonal pilot tone scheme, one of the pilot tones is used for estimating the polarization rotation angle, and the other one is used for compensating the polarization perturbation introduced phase noise. Meanwhile, residual effects are compensated precisely with the help of a real-valued FIR filter. In this case, the excess noise introduced by polarization perturbation is effectively suppressed. Experimental results show that the polarization scrambling rate ≥ 12.57 krad/s can be tracked by using the proposed algorithm, and a good estimated parameters performance is achieved. Compared to conventional polarization tracking algorithms such as the constant modulus algorithm (CMA), experimental results show that the polarization tracking capability of the proposed algorithm is significantly improved. Furthermore, much faster polarization tracking performance is evaluated by digital simulations, and the simulation results show that about 188.50 Mrad/s can be tracked by the proposed algorithm. Thus, our method provides effective technology for the practical application of fiber-based CV-QKD.

## I. INTRODUCTION

SECURE communication is a critical topic for next-generation fiber-optical systems, especially for communication networks with high-security requirements such as governmental, military, banking, or industry private networks. With the rapid development of advanced computing strategies, the traditional encryption technology that mainly depends on computational complexity becomes hard to guarantee the security of high-security required fiber-optical communication systems. The continuous-variable quantum key distribution (CV-QKD) with information theoretical security, which is compatible with optical communication networks, becomes a prime candidate to deal with security threats [1-2]. In this case, a lot of protocols and schemes have been proposed, among which the local local oscillator (LLO) CV-QKD with Gaussian modulated or discrete modulated coherent state protocol is a promising scheme for practical application [3-8]. For LLO CV-QKD, an independent free-running laser is used as the local oscillator, which has uncorrelated phase, wavelength and polarization compared to the laser at transmitter. Moreover, due to the random birefringence of the standard single-mode fiber (SSMF), the variation of the state of polarization (SOP) is drastic and random for quantum signals, and the rapid fluctuations of the SOP will significantly deteriorate the performance of the CV-QKD system [9-10]. Thus, the SOP of LLO CV-QKD systems needs to be manipulated to obtain good system performance.

Because of the low power of quantum signals, the SOP is mainly controlled by pilot tone for LLO CV-QKD systems, and optical or electrical methods are typically used for polarization tracking. For optical methods, SOP is tracked by a dynamic polarization controller with a feedback control circuit or manual polarization controller [11-13]. Since the SOP is tracked before coherent detection, the detection of quantum signals is not affected by high-power pilot tone signals. However, the channel transmittance will be decreased due to the insertion loss of SOP tracking devices, and it is also not beneficial for system integration. Furthermore, due to the limited speed of the polarization controller, it is not good to deal with a fast or burst variation of SOP. For electrical methods, the SOP launched into the receiver is random, and the quantum signal is demultiplexed by the digital signal processing (DSP) algorithms. Eriksson et al. [14] have demonstrated a CV-QKD system with fully digital phase and polarization tracking, where a mutation constant modulus algorithm (CMA) is used for polarization tracking. In Ref. [15], Kalman filter-based polarization tracking scheme is

Manuscript received xxx; revised xxx; accepted xxx. This work was supported by the National Key Research and Development Program of China (2020YFA0309704), the National Natural Science Foundation of China (U19A2076, U22A201034, 62101516, 62171418, 62201530 and 61901425), Basic Research Program of China (JCKY2021210B059), the Sichuan Science and Technology Program (2022ZYD0118, 2023JDRC0017, 2023YFG0143, 2022YFG0330, 2022ZDZX0009 and 2021YJ0313), the Natural Science Foundation of Sichuan Province (2023NSFSC1387 and 2023NSFSC0449), the Chengdu Major Science and Technology Innovation Program (2021-YF08-00040-GX), the Equipment Advance Research Field Foundation(315067206), the Chengdu Key Research and Development Support Program (2021-YF05-02430-GX and 2021-YF09-00116-GX), the Foundation of Science and Technology on Communication Security Laboratory (61421030402012111). *(Corresponding authors: Wei Huang, and Bingjie Xu).*

Yan Pan, Heng Wang, Yun Shao, Yaodi Pi, Ting Ye, Shuai Zhang, Yang Li, Wei Huang, and Bingjie Xu are with the Science and Technology on Communication Security Laboratory, Institute of Southwestern Communication, Chengdu, 610041, China (e-mail: pany14035@cetcsc.com; wangh7934@cetcsc.com; shaoyun@pku.edu.cn; piyd7836@cetcsc.com; yet17895@cetcsc.com; zhangshuai@cetcsc.com; yishuihanly@pku.edu.cn; huangw4548@cetcsc.com; xbjpku@pku.edu.cn).

employed to deal with SOP variation of the quantum signal in CV-QKD system, and simulations for different scrambling rates are performed to evaluate the performance of the algorithm, and about 1 krad/s SOP rotation can be resisted. In our previous work [16], a Stokes space-based polarization tracking algorithm is used for CV-QKD, and SOP tracking rate of $\geq 3$ krad/s is experimentally investigated. However, the complexity of above-mentioned algorithms is high, and the demonstrated polarization tracking performance for CV-QKD is limited to krad/s.

In this work, a simple and fast polarization tracking algorithm is proposed for LLO CV-QKD system. The algorithm is implemented by an orthogonal pilot tone scheme, one of the pilot tones is used for estimating the polarization rotation angle, and the other one is used for compensating the polarization perturbation introduced phase noise. Instead of multiple iterations, the proposed algorithm calculates the inverse Jones matrix by pilot tone directly. Therefore, compared to CMA, Kalman filter and Stokes space-based polarization demultiplexing algorithms, the complexity of the proposed algorithm is low. Moreover, residual effects are compensated precisely with the help of a real-valued FIR filter. Experimental results of 1 GBaud pilot-tone-assisted discrete modulated LLO CV-QKD system with a 24.49 km SSMF link are given, and the results of excess noise and secret key rates (SKRs) show that the proposed algorithm split the pilot-tone and quantum signal effectively. To evaluate polarization tracking performance, experiments under various scrambling rates are performed. The experimental results show that the proposed algorithm can track SOP scrambling of $\geq 12.57$ krad/s with a limited decrease of SKR. Compared to CMA, experimental results show that the polarization tracking capability of the proposed algorithm is significantly improved. Furthermore, the limiting polarization tracking performance of the proposed algorithm is evaluated by digital simulations, and the results show that the polarization tracking speed of the proposed algorithm can be reached about 188.50 Mrad/s. Therefore, the proposed algorithm can improve the flexibility and stability of LLO CV-QKD with low computing resources.

## II. OPERATING PRINCIPLE

The principle of the proposed algorithm is shown in Fig.1. At the transmitter, the quantum signal with discrete modulated coherent-state (DMCS) and pilot tone 1 (PT1) are modulated at vertical polarization, and the center frequency of PT1 is $f_{PT1}$. Here, the discrete Gaussian (DG) 256QAM modulation CV-QKD protocol is used under the security framework of Ref. [17]. The pilot tone 2 (PT2) is modulated at horizontal polarization, and the center frequency of the carrier is $f_{PT2}$. In this case, polarization and frequency division multiplexing are used to maximize the isolation between quantum and pilot tone signals, and denote the Jones vector of the multiplexed signal as $\left[ E_Q^{Tx}(t) + E_{PT1}^{Tx}(t) + \xi_V^{Tx}(t); E_{PT2}^{Tx}(t) + \xi_H^{Tx}(t) \right]^T$, where $E_Q^{Tx}(t)$ is the quantum signal, $E_{PT1}^{Tx}(t)$ is PT1, $\xi_V^{Tx}(t)$ is the excess noise at vertical polarization, $E_{PT2}^{Tx}(t)$ is PT2, and $\xi_H^{Tx}(t)$ is the excess noise at horizontal polarization. Then, the multiplexed signal is launched into the optical fiber link. Assuming the polarization characteristics of the optical fiber link can be represented by a unitary Jones matrix $J$ as described in Ref. [18], the received optical signal is

$$\begin{bmatrix} E_V^{Rx}(t) \\ E_H^{Rx}(t) \end{bmatrix} = J \times \begin{bmatrix} E_Q^{Tx}(t) + E_{PT1}^{Tx}(t) + \xi_V^{Tx}(t) \\ E_{PT2}^{Tx}(t) + \xi_H^{Tx}(t) \end{bmatrix},$$
$$\text{where } J = \begin{bmatrix} \cos\alpha(t)\exp(j\varphi_1(t)) & -\sin\alpha(t)\exp(j\varphi_2(t)) \\ \sin\alpha(t)\exp(-j\varphi_2(t)) & \cos\alpha(t)\exp(-j\varphi_1(t)) \end{bmatrix} \quad (1)$$

where $\alpha(t)$ is the polarization rotation angle, $\varphi_1(t)$ and $\varphi_2(t)$ are the phase angles introduced by polarization perturbation. Assuming the signal is detected by a polarization diversity receiver with identical delay and gain, signals with intermediate frequency are detected, which can be represented by

$$\begin{aligned} I_V(t) &= \cos\alpha(t)A_Q(t)\cos(2\pi f_{IF}t + \varphi_Q(t) + \varphi_n(t) + \varphi_1(t)) \\ &+ \cos\alpha(t)A_{PT1}(t)\cos(2\pi f_{PT1}t + \varphi_n(t) + \varphi_1(t)) \\ &- \sin\alpha(t)A_{PT2}(t)\cos(2\pi f_{PT2}t + \varphi_n(t) + \varphi_2(t)) + \xi_V^{Rx}(t) \\ I_H(t) &= \sin\alpha(t)A_Q(t)\cos(2\pi f_{IF}t + \varphi_Q(t) + \varphi_n(t) - \varphi_2(t)) \\ &+ \sin\alpha(t)A_{PT1}(t)\cos(2\pi f_{PT1}t + \varphi_n(t) - \varphi_2(t)) \\ &+ \cos\alpha(t)A_{PT2}(t)\cos(2\pi f_{PT2}t + \varphi_n(t) - \varphi_1(t)) + \xi_H^{Rx}(t) \end{aligned} \quad (2)$$

where $A_Q(t)$, $A_{PT1}(t)$ and $A_{PT2}(t)$ are the detected amplitude of quantum and pilot tone signals, $f_{IF}$, $f_{PT1}$ and $f_{PT2}$ are the intermediate frequency of quantum and pilot tone signals, $\varphi_Q(t)$ is the modulated phase of the quantum signal, $\varphi_n(t)$ is the phase noise introduced by laser's linewidth and fiber link scattering effects, etc. $\xi_V^{Rx}(t)$ and $\xi_H^{Rx}(t)$ are the excess noises at vertical and horizontal polarization after detection, respectively. Then, the DSP is performed to recover the raw key.

The DSP mainly includes: 1) Bandpass filtering. A frequency-domain ideal bandpass filter is used to split quantum, PT1 and PT2 signals. 2) Digital $x/p$ demodulation. The frequency estimation, down-conversion and lowpass filtering are performed to obtain $x$ and $p$ quadrature from the intermediate frequency signal digitally. Then, the Jones vector of quantum and pilot tone signals can be expressed by

$$\begin{bmatrix} E_Q^V(t) \\ E_Q^H(t) \end{bmatrix} = C_Q \begin{bmatrix} \cos\alpha(t)A_Q(t)\exp(j(\varphi_Q(t)+\varphi_n(t)+\varphi_1(t))) \\ \sin\alpha(t)A_Q(t)\exp(j(\varphi_Q(t)+\varphi_n(t)-\varphi_2(t))) \end{bmatrix} + \begin{bmatrix} \xi_Q^V(t) \\ \xi_Q^H(t) \end{bmatrix},$$
$$\begin{bmatrix} E_{PT1}^V(t) \\ E_{PT1}^H(t) \end{bmatrix} = C_{PT1} \begin{bmatrix} \cos\alpha(t)A_{PT1}(t)\exp(j(\varphi_n(t)+\varphi_1(t))) \\ \sin\alpha(t)A_{PT1}(t)\exp(j(\varphi_n(t)-\varphi_2(t))) \end{bmatrix} + \begin{bmatrix} \xi_{PT1}^V(t) \\ \xi_{PT1}^H(t) \end{bmatrix},$$
$$\begin{bmatrix} E_{PT2}^V(t) \\ E_{PT2}^H(t) \end{bmatrix} = C_{PT2} \begin{bmatrix} -\sin\alpha(t)A_{PT2}(t)\exp(j(\varphi_n(t)+\varphi_2(t))) \\ \cos\alpha(t)A_{PT2}(t)\exp(j(\varphi_n(t)-\varphi_1(t))) \end{bmatrix} + \begin{bmatrix} \xi_{PT2}^V(t) \\ \xi_{PT2}^H(t) \end{bmatrix}$$
(3)

where $C_Q$, $C_{PT1}$ and $C_{PT2}$ are the coefficients of digital demodulation for quantum and pilot tone signals, respectively. 3) Proposed algorithm. As can be seen from Eq. (3), the key to polarization tracking is how to calculate $\alpha(t)$, $\varphi_1(t)$ and $\varphi_2(t)$. Here, the following model is performed to demultiplexing the quantum and pilot signal

$$\begin{bmatrix} E^V(t) \\ E^H(t) \end{bmatrix} = J^{-1} \times \begin{bmatrix} E_{Q,PT1,PT2}^V(t) \\ E_{Q,PT1,PT2}^H(t) \end{bmatrix},$$

$$\text{where } J^{-1} = \begin{bmatrix} \cos\alpha(t)\exp(j\Delta\varphi(t)/2) & \sin\alpha(t)\exp(-j\Delta\varphi(t)/2) \\ -\sin\alpha(t)\exp(j\Delta\varphi(t)/2) & \cos\alpha(t)\exp(-j\Delta\varphi(t)/2) \end{bmatrix} \quad (4)$$

where $\Delta\varphi(t)$=-$[\varphi_1(t)+\varphi_2(t)]/2$. Using PT2 in Eq. (3), $\Delta\varphi(t)$ can be achieved by calculating the angle of $E_{PT2}^H(t) \bullet (E_{PT2}^V(t))^*$, and $\alpha(t)$ can be achieved by $\arctan(-E_{PT2}^V(t)/E_{PT2}^H(t) \bullet \exp(j\Delta\varphi(t)))$. Notably, the achieved angle should be unwrapped to avoid phase ambiguity. In this case, the inverse matrix $J^{-1}$ can be calculated, and the quantum and pilot tone signal are demultiplexed as

$$E_Q(t) = C_Q \begin{bmatrix} A_Q(t)\exp(j(\varphi_Q(t)+\varphi_n(t)+\phi(t))) + \zeta_Q^V(t) \\ \zeta_Q^H(t) \end{bmatrix},$$

$$E_{PT1}(t) = C_{PT1} \begin{bmatrix} A_{PT1}(t)\exp(j(\varphi_n(t)+\phi(t))) + \zeta_{PT1}^V(t) \\ \zeta_{PT1}^H(t) \end{bmatrix}, \quad (5)$$

$$E_{PT2}(t) = C_{PT2} \begin{bmatrix} \zeta_{PT2}^V(t) \\ A_{PT2}(t)\exp(j(\varphi_n(t)-\phi(t))) + \zeta_{PT2}^H(t) \end{bmatrix}$$

where $\zeta_{\{Q,PT1,PT2\}}^{\{V,H\}}(t)$ are the noises, and $\phi(t)$=$[\varphi_1(t)$-$\varphi_2(t)]/2$. 4) Phase noise compensation. The phase noise $\varphi_n(t)$ and $\phi(t)$ are estimated from pilot tone signals in Eq. (5), and the phase noise of quantum signals can be compensated. 5) Data-aided equalization. As can be seen from Eq. (5), the quantum and pilot signals can be split in the dimension of polarization theoretically. However, because of the nonideal estimation of $\alpha(t)$ and $\Delta\varphi(t)$, polarization variation introduced noise cannot be completely suppressed. Here, with the help of the training sequence and least-mean-square (LMS) algorithm, a real-valued FIR filter is designed to compensate for the residual noise [19]. After the above processing, the raw key signal $\hat{E}_Q^{Tx}(t)$ is obtained.

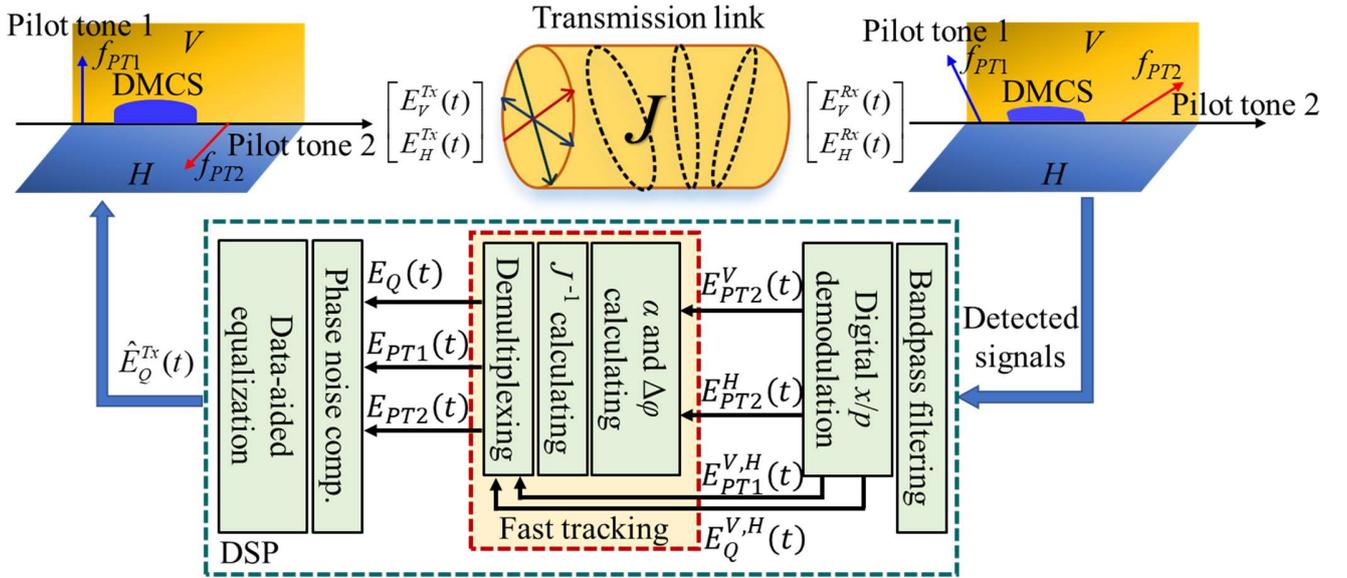

**Fig. 1.** The principle of orthogonal pilot-tone-assisted fast polarization tracking algorithm. comp., compensation.

### III. EXPERIMENTAL SETUP

Figure 2 shows the experimental setup of the LLO CV-QKD system to evaluate the performance of the proposed algorithm. At Alice's site, a continuous-wave laser (i.e., CW laser 1) with a linewidth of <100 Hz is used as the carrier, and the wavelength is set to 1550.22 nm. Then, the light is split into two branches by a polarization beam splitter (PBS). One branch of the optical carrier is modulated by an In-phase/quadrature modulator (IQ modulator). The $x$ and $p$ quadrature signals (i.e., DG-256QAM as shown in Ref. [20]) with an 850 MHz frequency shift are generated by a two channels arbitrary waveform generator (AWG) that works at 30 GSa/s, and the two electrical signals are amplified by two amplifiers for driving the IQ modulator. Here, PT1 with 850 MHz frequency shift from the center frequency of the quantum signal is modulated, and the symbol rate of the quantum signal is set to 1 GBaud. Then, a variable optical attenuator (VOA) is used to adjust the modulation variance $V_A$, and the optical DMCS quantum signal can be obtained. The state of polarization (SOP) of the optical DMCS signal is controlled by a polarization controller (PC) and aligns with the principal axis of the polarization beam combiner (PBC). The reference path mainly consists of an optical delay line (ensure the reference signal has a similar phase characteristic to the optical DMCS signal), VOA (control the optical power of the reference signal), and PC (align the SOP of the reference signal to the other principal axis of PBC). Therefore, the optical DMCS quantum signal and high-power reference signal are multiplexed by the dimension of polarization and frequency. In the transmission link, 24.49 km SSMF is used in the laboratory environment. Before being launched into the receiver, the SOP of the optical signal is continuously scrambled by General Photonic PSY-201. At Bob's site, another independent running CW laser with a linewidth of <100 Hz is used as the local oscillator, and the center wavelength is about 1.75 GHz shift from CW laser 1. Then, the optical signal and local oscillator are coherently detected by a polarization diversity receiver. The polarization diversity receiver consists of two PBS, two polarization-maintaining optical couplers (PMOC), and two balanced photodetectors (BPD). The 3-dB bandwidth, responsibility, and gain of the BPD are 1.6 GHz, 0.85 $A/W$, and $1.6×10^4$ $V/A$, respectively. Finally, the received electrical signals are digitalized by a digital storage oscilloscope (DSO) at 10 GSa/s, and offline DSP shown in Fig.1 is performed for raw key recovery.

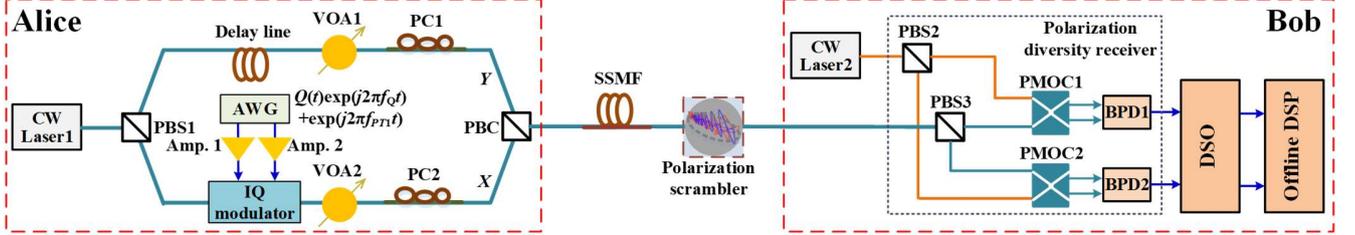

**Fig. 2.** The experimental setup of LLO CV-QKD to evaluate the performance of the proposed algorithm.

## IV. EXPERIMENTAL RESULTS AND DISCUSSION

Figure 3 shows the experimental results of measured excess noise and calculated SKRs at different polarization scrambling rates (SRs) to evaluate the polarization tracking performance of the proposed algorithm. Here, 20 tests have been performed at each SR, and the excess noise is estimated with a block size of $1×10^6$. Fig. 3(a) shows the excess noise performance under different SRs, and the fluctuation of excess noise is around 0.02-0.06 in the shot noise unit (SNU). The pink stars are the mean value of excess noises, which are 0.038, 0.038, 0.043, 0.052, and 0.048 SNU under SR of 0.63, 1.26, 3.14, 6.28 and 12.57 krad/s, respectively. Fig. 3(b) shows the excess noise performance under SR=12.57 krad/s (i.e., the fastest SR of PSY-201) when the fast polarization tracking algorithm is used or not, and the results of SR=0 krad/s are given as a reference. As can be seen from Fig. 3(b), the fluctuation of excess noise for SR=0 and 12.57 krad/s are similar when fast polarization tracking algorithm is used, and the excess noise is unacceptable without fast polarization tracking. Fig. 3(c) shows the experimental results of achieved SKR under different SRs. The curve represents the asymptotic SKR bound by utilizing the parameters of SR=0 krad/s where the modulation variance, excess noise, and loss of fiber link are 6.15 SNU, 0.030 SNU, and 4.971dB, respectively. Furthermore, the quantum efficiency of the receiver $η$=0.56, the electronic noise $V_{ele}$=0.15 SNU, reconciliation efficiency $β$=0.95, and the ratio of the training sequence is set to 20%. Dots in Fig. 3(c) represent the experimental measured average SKRs, which are calculated by a similar method in Ref. [20]. From Fig. 3(c), it is hard to distinguish the SKRs for different SRs. The asymptotic SKRs are 51.60, 45.98, 46.66, 43.82, 39.56, and 42.14 Mbps at SR of 0, 0.63, 1.26, 3.14, 6.28, 12.57 krad/s, respectively. In this case, the proposed fast polarization tracking algorithm is effective for correcting fast or bursting SOP perturbation.

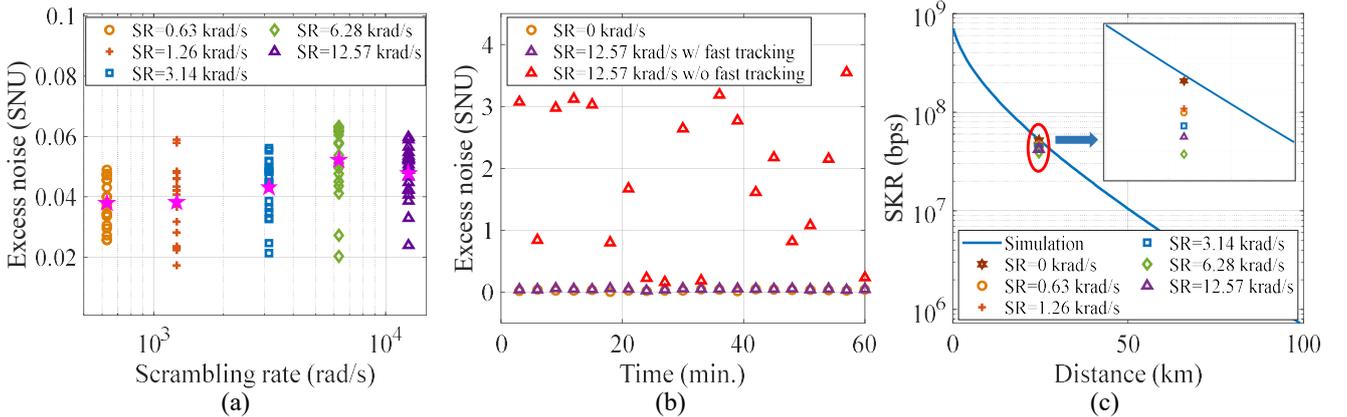

**Fig. 3.** The polarization tracking performance of the proposed algorithm for LLO CV-QKD. (a) the excess noise performance at different SRs, (b) the excess noise performance at SR=12.57 krad/s with and without fast polarization tracking algorithm, and (c) the SKR performance at different SRs.

To further demonstrate the polarization tracking performance of the proposed algorithm, comparisons with different algorithms at different SRs have shown in Fig. 4. Fig. 4(a) shows the excess noise performance by using data-aided real-valued FIR filter for polarization tracking, which is described in Ref. [19] in detail. The length of filter tap is set to 11, and the step-size of LMS is set to $10^{-5}$. Fig. 4(b) shows the excess noise performance by using CMA for polarization tracking, and 1-tap polarization demultiplexing stage with a fixed step-size of $10^{-5}$ is used. Fig. 4(c) shows the average SKR performance comparison by using a data-aided real-valued FIR filter, CMA and proposed algorithm for polarization tracking. As can be seen from Fig. 4(a) and (b), not like the results shown in Fig. 3(a), the excess noises are extremely pessimistic when SR=6.28 krad/s and 12.57 krad/s. Furthermore, the comparison of average SKR can better present the polarization tracking advantages of the proposed algorithm. From Fig. 4(c), we can find that the average SKR performance degradation of the proposed algorithm is much lower compared with the data-aided real-valued FIR filter and CMA.

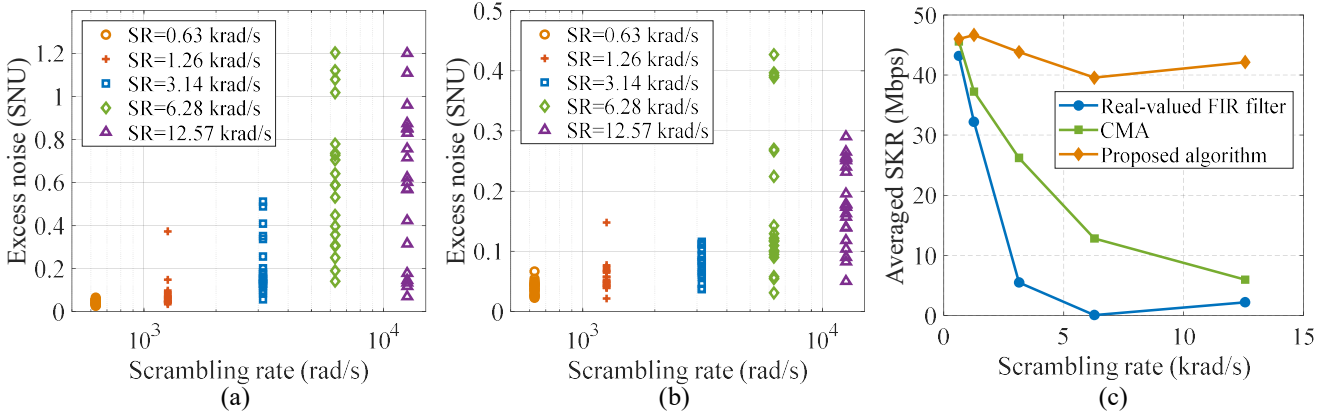

**Fig. 4.** The polarization tracking performance comparison of different algorithms at different SRs. (a) The excess noise performance by using a real-valued FIR filter for polarization tracking, (b) the excess noise performance by using CMA for polarization tracking, and (c) the averaged SKR performance comparison by using real-valued FIR filter, CMA and proposed algorithm for polarization tracking.

Obviously, due to our device's limitation, this level of scrambling rate is insufficient to show the ability of the proposed algorithm. Hence, based on the experimental data, an equivalent polarization rotation matrix is used to simulate higher-speed polarization scrambling in the following analysis, and the simulation results are shown in Fig. 5. Notably, the polarization rotation matrix is $J = [\cos(kwT_s) \sin(kwT_s); -\sin(kwT_s) \cos(kwT_s)]$, where $w$ represents the polarization rotation angular frequency, and $T_s$ is the symbol period [21]. In Fig. 5, the left Y-axe shows the excess noise performance versus different scrambling rates, and the right Y-axe shows the average SKR performance versus different scrambling rates. As can be seen from Fig. 5, the excess noise and average SKR performance do not decrease significantly even when SR≤188.50 Mrad/s. Thus, the proposed algorithm has an excellent polarization tracking performance for the studied LLO CV-QKD system.

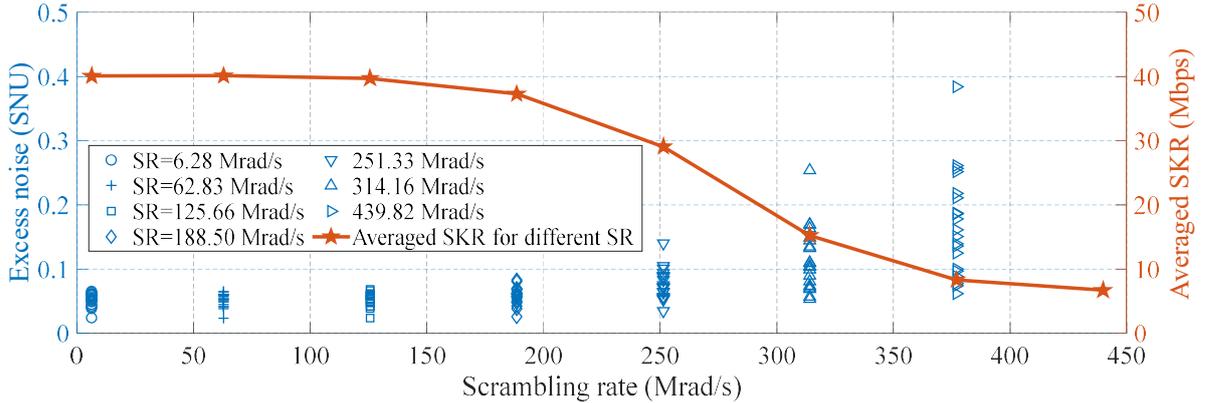

**Fig. 5.** Simulation results of the proposed algorithm for different SR.

## V. CONCLUSION

A simple and fast polarization tracking algorithm is proposed and experimentally demonstrated for LLO CV-QKD system. The results of excess noise and SKR for 1 Gbaud DG-256QAM after transmission of 24.49 km SSMF are given, and it shows the proposed algorithm has a good performance for the LLO CV-QKD system. More importantly, the algorithm performance is tested under different SRs, which is a big challenge for demultiplexing the quantum signals. The results show that the performance of excess noise and SKR is remarkable when the SR is 12.57 krad/s. Meanwhile, the polarization tracking performance comparison of different algorithms is given, which show that the proposed algorithm has a more stable excess noise and average SKR performance compared with data-aided real-valued FIR filter and CMA.

Furthermore, to evaluate the performance of the proposed algorithm, digital simulations with SR up to 439.82 Mrad/s are performed, and the simulation results show that the excess noise and average SKR performance do not decrease significantly when SR≤188.50 Mrad/s. Consequently, the proposed fast polarization tracking algorithm has the potential for improving the flexibility and stability of LLO CV-QKD.